\documentclass[twocolumn,showpacs,amsmath,amssymb,aps,prd]{revtex4}

\usepackage{bm}

\newcommand{\itGamma}{{\mathit{\Gamma}}}
\newcommand{\del}{\partial}

\newcommand{\cM}{{\cal M}}
\newcommand{\cO}{{\cal O}}
\newcommand{\cR}{{\cal R}}
\newcommand{\cT}{{\cal T}}
\newcommand{\ds}{\displaystyle}

\newcommand{\lc}{\varepsilon}

\newcommand{\diag}{\mathop{\rm diag}\nolimits}
\newcommand{\orto}{{\scriptscriptstyle\perp}}
\newcommand{\para}{\scriptscriptstyle\parallel}
\newcommand{\Pp}{ {P_{\para}}\vphantom{P} }
\newcommand{\Pn}{ {P_{\orto}}\vphantom{P} }
\newcommand{\Ppp}{ {p_{\para}}\vphantom{p} }
\newcommand{\Pnn}{ {p_{\orto}}\vphantom{p} }
\newcommand{\cfl}[2]{{\textstyle {{#1}\brace {#2}}}}
\newcommand{\h}{h}

\begin{document}

\title{Spinning branes in Riemann-Cartan spacetime}

\author{Milovan Vasili\'c}
 \email{mvasilic@phy.bg.ac.yu}
\author{Marko Vojinovi\'c}
 \email{vmarko@phy.bg.ac.yu}
\affiliation{Institute of Physics, P.O.Box 57, 11001 Belgrade, Serbia}

\date{\today}

\begin{abstract}
We use the conservation law of the stress-energy and spin tensors to
study the motion of massive brane-like objects in Riemann-Cartan geometry. The
world-sheet equations and boundary conditions are obtained in a manifestly
covariant form. In the particle case, the resultant world-line equations turn out
to exhibit a novel spin-curvature coupling. In particular, the spin of a
zero-size particle does not couple to the background curvature. In the string
case, the world-sheet dynamics is studied for some special choices of spin and
torsion. As a result, the known coupling to the Kalb-Ramond antisymmetric
external field is obtained. Geometrically, the Kalb-Ramond field has been
recognized as a part of the torsion itself, rather than the torsion potential.
\end{abstract}


\maketitle

\section{\label{IntroductionSection}Introduction}

The problem of motion of brane like objects in backgrounds of nontrivial
geometry is addressed by using some form of the Mathisson-Papapetrou method
\cite{Mathisson1937, Papapetrou1951}. One starts with the covariant
conservation law of the stress-energy and spin tensors of matter fields, and
analyzes it under the assumption that matter is localized to resemble a
brane. In the lowest, single-pole approximation, the moving matter is viewed
as an infinitely thin brane. In the pole-dipole approximation, its non-zero
thickness is taken into account.

The known results concerning extended objects in Riemann-Cartan geometry
exclusively refer to particles. They can be summarized as follows. Spinless
particles in the single-pole approximation obey the geodesic equation. In the
pole-dipole approximation, the rotational angular momentum of the localized
matter couples to spacetime curvature, and produces geodesic deviation
\cite{Mathisson1937, Papapetrou1951, Tulczyjew1959, Taub1964, Dixon}. If the particles have spin, the curvature couples to the
total angular momentum, and the torsion to the spin alone \cite{Trautman1972,
Hehl1976a, Yasskin1980, Nomura1991, Nomura1992}.

As for the higher branes, the results found in literature exclusively refer
to spinless matter and Riemannian backgrounds \cite{Vasilic2006,
Vasilic2007}. It has been shown that spacetime curvature couples to the
internal angular momentum of a thick brane, and that this coupling
disappears if the brane is infinitely thin.

In this paper, we shall study spinning brane-like matter in spacetimes with
curvature and torsion. Our motivation is threefold. First, realistic strings
(like flux tubes) are really believed to exist, and to be relevant for the
description of hadronic matter. Second, we want to check if the presence of
matter with spin saves the spin-curvature coupling even if the brane has no
thickness. Finally, the influence of torsion on the brane dynamics can
provide a geometric insight into the extended string actions found in
literature. Namely, the basic Nambu-Goto string action \cite{Nambu, Goto} is
in literature often modified to include interaction with additional
background fields. Apart from the target-space metric, the antisymmetric
tensor field ${\cal B}_{\mu\nu}(x)$ and the dilaton field $\Phi(x)$ are
considered \cite{B1, B2, B3, B4}. While spacetime metric has obvious
geometric interpretation, the background fields ${\cal B}_{\mu\nu}(x)$ and
$\Phi(x)$ do not. The attempts have been made in literature to interpret
${\cal B}_{\mu\nu}$ and $\Phi$ as originating from the background torsion
and non-metricity, respectively \cite{G1, G2, G3, G4, G5, Nepomechie1985,
Freund1982}. Our idea is to consider stringy shaped matter in backgrounds of
general geometry, and check if the effective action of Refs. \cite{B1, B2,
B3, B4} could be recovered. This way, the real geometric nature of the
background fields ${\cal B}_{\mu\nu}(x)$ and $\Phi(x)$ could be found.

The results that we have obtained are summarized as follows. The world-sheet
equations and boundary conditions for a $p$-dimensional brane in $D$-dimensional
Riemann-Cartan spacetime are derived in a manifestly covariant way. It has been
shown that spacetime curvature couples to $(p+1)$-dimensional currents associated
with the internal angular momentum of the brane, while torsion couples to the
spin alone. The curvature coupling disappears in the limit of an infinitely thin
brane, in spite of the fact that brane is made of spinning matter. As
illustrative examples, the $0$-brane (particle) and $1$-brane (string) are given
an additional consideration. The particle dynamics has been found to differ from
what has been believed so far. In particular, the spin of a strict point particle
does not couple to the curvature. We have also analyzed the world-sheet equations
and boundary conditions of an infinitely thin string. The generalized string
action of Refs. \cite{B1, B2, B3, B4} has been recovered by assigning a
special value to the spin tensor and the background torsion. According to our
results, the Kalb-Ramond antisymmetric tensor field ${\cal B}_{\mu\nu}(x)$ is related
to the torsion itself, rather than to its potential as suggested in literature.

The layout of the paper is as follows. In section \ref{Formalism}, we define
the conservation law of the stress-energy and spin tensors, and introduce
the necessary geometric notions. Using the fact that antisymmetric part
of the stress-energy tensor is completely determined by the spin tensor, we
eliminate it from further considerations. The conservation equations are
rewritten in terms of the independent variables---the spin tensor and the
generalized Belinfante tensor. After the brief recapitulation of the
covariant multipole formalism, we define the pole-dipole approximation for
the independent variables, only. Section \ref{EOMSection} is devoted to the
derivation of the brane world-sheet equations. The actual derivation is only
sketched, as the method has already been analyzed in detail in
\cite{Vasilic2007}. The world-sheet equations and boundary conditions are
obtained in a manifestly covariant form. In Section \ref{Examples},
particles and strings are given additional consideration. In the particle
case, the resulting equations of motion are compared to the pole-dipole
equations found in literature \cite{Yasskin1980, Nomura1991}. As it turns
out, they coincide in the pole-dipole approximation, but have different
single-pole limits. In the string case, the world-sheet equations are
analyzed in the zero-thickness limit. By an appropriate choice of the
spin-tensor and the background torsion, we have recovered the effective
dynamics of Refs. \cite{B1, B2, B3, B4}. In section \ref{ConclusionSection},
we give our final remarks.

Our conventions are the same as in Ref.\ \cite{Vasilic2007}. Greek indices
$\mu,\nu,\dots$ are the spacetime indices, and run over $0,1,\dots,D-1$. Latin
indices $a,b,\dots$ are the world-sheet indices and run over $0,1,\dots,p$. The
Latin indices $i,j,\dots$ refer to the world-sheet boundary and take values
$0,1,\dots,p-1$. The coordinates of spacetime, world-sheet and world-sheet
boundary are denoted by $x^{\mu}$, $\xi^a$ and $\lambda^i$, respectively. The
corresponding metric tensors are denoted by $g_{\mu\nu}(x)$, $\gamma_{ab}(\xi)$
and $\h_{ij}(\lambda)$. The signature convention is defined by
$\diag(-,+,\dots,+)$, and the indices are raised by the inverse metrics
$g^{\mu\nu}$, $\gamma^{ab}$ and $\h^{ij}$.

\section{\label{Formalism}The multipole formalism}

We begin with the covariant conservation of the fundamental matter currents ---
stress-energy tensor $\tau^{\mu}{}_{\nu}$, and spin tensor
$\sigma^{\lambda}{}_{\mu\nu}$:
\begin{subequations} \label{ZakoniOdrzanja}
\begin{equation} \label{ZakonOdrzanjaTau}
\left( D_{\nu} + \cT^{\lambda}{}_{\nu\lambda} \right) \tau^{\nu}{}_{\mu} =
\tau^{\nu}{}_{\rho} \cT^{\rho}{}_{\mu\nu} + \frac{1}{2}
\sigma^{\nu\rho\sigma}\cR_{\rho\sigma\mu\nu},
\end{equation}
\begin{equation} \label{ZakonOdrzanjaSigma}
\left( D_{\nu} + \cT^{\lambda}{}_{\nu\lambda}
\right)\sigma^{\nu}{}_{\rho\sigma} = \tau_{\rho\sigma} - \tau_{\sigma\rho}.
\end{equation}
\end{subequations}
Here, $D_{\nu}$ is the covariant derivative with the nonsymmetric connection
$\itGamma^{\lambda}{}_{\mu\nu}$, which acts on a vector $v^{\mu}$ according
to the rule $D_{\nu} v^{\mu} \equiv \del_{\nu}v^{\mu} +
\itGamma^{\mu}{}_{\lambda\nu}v^{\lambda}$. The torsion
$\cT^{\lambda}{}_{\mu\nu}$, and curvature $\cR^{\mu}{}_{\nu\rho\sigma}$ are
defined in the standard way:
\begin{eqnarray}
& \cT^{\lambda}{}_{\mu\nu} \equiv \itGamma^{\lambda}{}_{\nu\mu} -
\itGamma^{\lambda}{}_{\mu\nu}, & \nonumber \\
& \cR^{\mu}{}_{\nu\rho\sigma} \equiv
\del_{\rho} \itGamma^{\mu}{}_{\nu\sigma} - \del_{\sigma}
\itGamma^{\mu}{}_{\nu\rho} + \itGamma^{\mu}{}_{\lambda\rho}
\itGamma^{\lambda}{}_{\nu\sigma} - \itGamma^{\mu}{}_{\lambda\sigma}
\itGamma^{\lambda}{}_{\nu\rho}. & \nonumber
\end{eqnarray}

The derivative $D_{\lambda}$ is assumed to satisfy the metricity condition,
$D_{\lambda}g_{\mu\nu}=0$. As a consequence, the connection
$\itGamma^{\lambda}{}_{\mu\nu}$ is split into the Levi-Civita connection
$\cfl{\lambda}{\mu\nu}$, and the contorsion $K^{\lambda}{}_{\mu\nu}$:
\begin{eqnarray}
& \itGamma^{\lambda}{}_{\mu\nu} = \cfl{\lambda}{\mu\nu} +
K^{\lambda}{}_{\mu\nu}, & \nonumber \\
& \ds K^{\lambda}{}_{\mu\nu} \equiv -\frac{1}{2}
\left( \cT^{\lambda}{}_{\mu\nu} - \cT_{\nu}{}^{\lambda}{}_{\mu} +
\cT_{\mu\nu}{}^{\lambda} \right). & \nonumber
\end{eqnarray}
We shall also introduce the Riemannian covariant derivative $\nabla_{\mu}
\equiv D_{\mu}(\itGamma\to \{ \} )$, and the Riemannian curvature tensor
$R^{\mu}{}_{\nu\rho\sigma} \equiv \cR^{\mu}{}_{\nu\rho\sigma}(\itGamma \to
\{ \} )$. The relation connecting the two curvature tensors reads:
$$
\cR^{\mu}{}_{\nu\lambda\rho} = R^{\mu}{}_{\nu\lambda\rho} + 2
\nabla_{[\lambda} K^{\mu}{}_{\nu\rho]} + 2 K^{\mu}{}_{\sigma[\lambda}
K^{\sigma}{}_{\nu\rho]},
$$
where the indices in square brackets are antisymmetrized.

Given the system of conservation equations (\ref{ZakoniOdrzanja}), one finds
that the second one has no dynamical content. Indeed, the antisymmetric part
of stress-energy tensor is completely determined by the spin tensor. One can
use (\ref{ZakonOdrzanjaSigma}) to eliminate $\tau^{[\mu\nu]}$ from the
equation (\ref{ZakonOdrzanjaTau}), and thus obtain the conservation
equation, in which only $\tau^{(\mu\nu)}$ and $\sigma^{\lambda\mu\nu}$
components appear. The resulting equation can be written in the form
\begin{eqnarray} \label{GlavniZakonOdrzanja}
 & \ds \nabla_{\nu}\left( \theta^{\mu\nu} -
K^{[\mu}{}_{\lambda\rho} \sigma^{\rho\lambda\nu]} -
\frac{1}{2} K_{\lambda\rho}{}^{[\mu}\sigma^{\nu]\rho\lambda}\right) = & \nonumber \\
 & \ds = \frac{1}{2}\sigma_{\nu\rho\lambda}\nabla^{\mu} K^{\rho\lambda\nu} \,, &
\end{eqnarray}
where $\theta^{\mu\nu}=\theta^{\nu\mu}$ stands for the generalized
Belinfante tensor:
\begin{equation} \label{Belinfante}
\theta^{\mu\nu}\equiv
\tau^{(\mu\nu)} - \nabla_{\rho}\sigma^{(\mu\nu)\rho} -
\frac{1}{2} K_{\lambda\rho}{}^{(\mu}\sigma^{\nu)\rho\lambda}\,.
\end{equation}
The independent variables $\theta^{\mu\nu}$ and $\sigma^{\mu\nu\rho}$ are in
1--1 correspondence with the original variables. In what follows, the
conservation law in the form (\ref{GlavniZakonOdrzanja}) will be the
starting point for the derivation of the brane world-sheet equations.

Let us now introduce the multipole formalism, which is necessary for the
derivation. It has been shown in Refs. \cite{Vasilic2006, Vasilic2007} that
an exponentially decreasing function can be expanded into a series of
$\delta$-function derivatives. For example, a tensor valued function
$F^{\mu\nu}(x)$, well localized around the $(p+1)$-dimensional surface $\cM$
in $D$-dimensional spacetime, can be decomposed in a manifestly covariant
way as
\begin{eqnarray} \label{DeltaSeries}
& \ds F^{\mu\nu}(x) = \int_{\cM} d^{p+1}\xi \sqrt{-\gamma}
\bigg[ M^{\mu\nu}\,\frac{\delta^{(D)}(x-z)}{\sqrt{-g}} - & \nonumber \\
& \ds - \nabla_{\rho}\left(M^{\mu\nu\rho}
\frac{\delta^{(D)}(x-z)}{\sqrt{-g}}\right) + \cdots \bigg]. &
\end{eqnarray}
The surface $\cM$ is defined by the equation $x^{\mu}=z^{\mu}(\xi)$, where
$\xi^a$ are the surface coordinates, and the coefficients $M^{\mu\nu}(\xi)$,
$M^{\mu\nu\rho}(\xi)$, ... are spacetime tensors called multipole
coefficients. Here, and in what follows, we shall frequently use the notion
of the surface coordinate vectors
$$
u_a^{\mu} = \frac{\del z^{\mu}}{\del \xi^a} \, ,
$$
and the surface induced metric tensor
$$
\gamma_{ab} = g_{\mu\nu} u_a^{\mu} u_b^{\nu} \, .
$$
The induced metric is assumed to be nondegenerate, $\gamma \equiv
\det(\gamma_{ab}) \neq 0$, and of Minkowski signature. The same holds for
the target space metric $g_{\mu\nu}(x)$ and its determinant $g(x)$.

It has been shown in Ref. \cite{Vasilic2007} that one may truncate the
series in a covariant way in order to approximate the description of matter.
Truncation after the leading term is called \emph{single-pole}
approximation, truncation after the second term is called \emph{pole-dipole}
approximation. The physical interpretation of these approximations is the
following. In the single-pole approximation, one assumes that the brane
has no thickness, which means that matter is localized on a surface. All
higher approximations, including pole-dipole, allow for the non-zero
thickness, and thus, for the transversal internal motion.

Apart from being covariant with respect to diffeomorphisms, the series
(\ref{DeltaSeries}) possesses two extra gauge symmetries. The first is a
consequence of the fact that that there are redundant coefficients in this
decomposition. Indeed, only $D-p-1$ out of $D$ $\delta$-functions in each
term of the multipole expansion (\ref{DeltaSeries}) are effective in
modeling trajectory of a brane like object in $D$-dimensional spacetime. The
$p+1$ extra $\delta$-functions and the extra integrations are introduced
only to covariantize the expressions. The derivatives parallel to the
world-sheet are integrated out, as they should, considering the fact that
matter is not localized along the brane. As a consequence, the parallel
components of the higher multipole coefficients are expected to effectively
disappear. It has been shown in Ref. \cite{Vasilic2007} that the
corresponding gauge symmetry, named \emph{extra symmetry}~1, in the
pole-dipole approximation has the form
\begin{subequations} \label{extra1}
\begin{equation} \label{extra1a}
\delta_1 M^{\mu\nu} = \nabla_a \epsilon^{\mu\nu a} \,, \qquad
\delta_1 M^{\mu\nu\rho} = \epsilon^{\mu\nu a}u^{\rho}_a \,,
\end{equation}
where $\epsilon^{\mu\nu a}(\xi)$ are gauge parameters constrained by the
boundary condition
\begin{equation} \label{extra1b}
n_a\epsilon^{\mu\nu a}|_{\partial\cM}=0 \,.
\end{equation}
\end{subequations}
Here, $n^a$ is the unit boundary normal, and $\nabla_a$ stands for the
world-sheet total covariant derivative (see the Appendix). Thus, the parallel
components of $M^{\mu\nu\rho}$ transform as $\delta_1
(u^a_{\rho}M^{\mu\nu\rho})=\epsilon^{\mu\nu a}\,$, and can be gauged away. In
fact, one can show that the parallel com\-po\-nents of the higher multipoles are
also pure gauge. In the gauge fixed multipole expansion, the only derivatives
that appear are those orthogonal to the world-sheet. In the single-pole
approximation, the extra symmetry 1 is trivial.

The second extra symmetry stems from the fact that the choice of the surface
$x^{\mu}=z^{\mu}(\xi)$ in the expansion (\ref{DeltaSeries}) is arbitrary. If
we use another surface, let us say $x^{\mu} = z'^{\mu}(\xi)$, the
coefficients $M^{\mu\nu}$, $M^{\mu\nu\rho}$, ... will change to
$M'^{\mu\nu}$, $M'^{\mu\nu\rho}$, ... while leaving the function
$F^{\mu\nu}(x)$ invariant. The transformation law of the $M$-coefficients,
generated by the replacement $z^{\mu}\to z'^{\mu}$, defines the gauge
symmetry that we call \emph{extra symmetry}~2.

The extra symmetry 2 is an exact symmetry of the full expansion
(\ref{DeltaSeries}), but only approximate symmetry of the truncated series.
In the pole-dipole approximation, it has the form
\begin{equation} \label{extra2}
\begin{array}{lcl}
\delta_2 z^{\mu} & = & \epsilon^{\mu}\,,                       \\
\delta_2 M^{\mu\nu} & = & - M^{\mu\nu} u^a_{\rho} \nabla_a \epsilon^{\rho} -
M^{\lambda\nu} \cfl{\mu}{\lambda\rho}\epsilon^{\rho} -
M^{\mu\lambda} \cfl{\nu}{\lambda\rho}\epsilon^{\rho} ,       \\
\delta_2 M^{\mu\nu\rho} & = & - M^{\mu\nu}\epsilon^{\rho} \,,  \\
\end{array}
\end{equation}
provided the $M$-coefficients are subject to the hierarchy
$M^{\mu\nu}=\cO_0$, $M^{\mu\nu\rho}=\cO_1$, and the free parameters
$\epsilon^{\mu}(\xi)$ satisfy $\epsilon^{\mu}=\cO_1$. Here, $\cO_n$ stands
for the order of smallness, and the condition $\epsilon^{\mu} = \cO_1$
ensures that the order of truncation is not violated by the action of the
symmetry transformations \cite{Vasilic2007}. In the pole-dipole and higher
approximations, fixing the gauge of extra symmetry 2 defines the central
surface of mass. In the single-pole approximation, the extra symmetry 2 is
trivial.

Now, we shall replace the general function $F^{\mu\nu}(x)$ with the
stress-energy and spin tensors of the localized matter. In the pole-dipole
approximation, our independent variables $\theta^{\mu\nu}$ and
$\sigma^{\lambda\mu\nu}$ are written in the form
\begin{widetext}
\begin{subequations} \label{PoleDipole}
\begin{equation} \label{DeltaRazvojZaTau}
\theta^{\mu\nu} = \int d^{p+1}\xi \sqrt{-\gamma}
\left[ B^{\mu\nu}\,\frac{\delta^{(D)}(x-z)}{\sqrt{-g}} -
\nabla_{\rho}\left(B^{\mu\nu\rho}\frac{\delta^{(D)}(x-z)}{\sqrt{-g}}\right)
\right]\,,
\end{equation}
\begin{equation} \label{DeltaRazvojZaSigma}
\sigma^{\lambda\mu\nu} = \int d^{p+1}\xi \sqrt{-\gamma}\, C^{\lambda\mu\nu}
\frac{\delta^{(D)}(x-z)}{\sqrt{-g}}\,,
\end{equation}
\end{subequations}
\end{widetext}
where $B^{\mu\nu}(\xi)$, $B^{\mu\nu\rho}(\xi)$ and $C^{\lambda\mu\nu}(\xi)$
are the corresponding multipole coefficients. As we can see, the
decomposition of the spin-tensor $\sigma^{\lambda\mu\nu}$ lacks the dipole
term. This is because the spin is considered to be of the order of the
orbital angular momentum which is already described by the dipole
coefficient. Any correction to the leading term  of
(\ref{DeltaRazvojZaSigma}) would, therefore, take us beyond the pole-dipole
approximation.

The symmetries of the expansion (\ref{PoleDipole}) are basically the same as
found in the general considerations. They include spacetime and world-sheet
diffeomorphisms, and two extra symmetries. The transformation properties of
the multipole coefficients with respect to spacetime diffeomorphisms and
world-sheet reparametrizations are determined by their index structure.
Thus, $B^{\mu\nu}$, $B^{\mu\nu\rho}$ and $C^{\lambda\mu\nu}$ are spacetime
tensors, and world-sheet scalars. As for the two extra symmetries, the
transformation law of the $B$-coefficients is given by the general formulas
(\ref{extra1}) and (\ref{extra2}), but the $C$-coefficients transform
trivially:
\begin{equation} \label{extraC}
\delta_1 C^{\lambda\mu\nu}=\delta_2 C^{\lambda\mu\nu}=0\,.
\end{equation}
This is because the expansion (\ref{DeltaRazvojZaSigma}) has a single-pole
form, and we have already established that extra symmetries in the
single-pole approximation are trivial. In addition, the multiple
coefficients are required to obey the hierarchy $B^{\mu\nu}=\cO_0$,
$B^{\mu\nu\rho}\sim C^{\lambda\mu\nu}=\cO_1$. Only then the extra symmetry 2
remains unbroken.

In what follows, we shall use the expansion (\ref{PoleDipole}) to solve the
conservation equations (\ref{GlavniZakonOdrzanja}), and define the limit of an
infinitely thin brane.

\section{\label{EOMSection}Equations of motion}

In this section, the stress-energy and spin-tensor conservation equations
(\ref{GlavniZakonOdrzanja}) are analyzed in the pole-dipole approximation.
The brane world-sheet equations and boundary conditions are derived in a
manifestly covariant way.

\subsection{\label{Derivation}Derivation}

The brane world-sheet equations are derived in the following way. We insert
(\ref{PoleDipole}) into (\ref{GlavniZakonOdrzanja}), and solve for
the unknown variables $z^{\mu}(\xi)$, $B^{\mu\nu}(\xi)$, $B^{\mu\nu\rho}(\xi)$
and $C^{\lambda\mu\nu}(\xi)$. The algorithm for solving this type of equation is
discussed in detail in \cite{Vasilic2006, Vasilic2007}, and here we only sketch
it. The first step is to multiply the equation (\ref{GlavniZakonOdrzanja}) with
an arbitrary spacetime function $f_{\mu}(x)$ of compact support, and integrate
over the spacetime. The resulting equation depends on the function $f_{\mu}$ and
its first and second covariant derivatives, evaluated on the surface
$x^{\mu}=z^{\mu}(\xi)$. Precisely, we obtain
\begin{eqnarray} \label{jna20}
& \ds \int d^{p+1}\xi \sqrt{-\gamma} \,
\bigg[ B^{\mu\nu\rho} f_{\mu;\nu\rho} +
\left(B^{\mu\nu}-D^{\mu\nu}\right) f_{\mu;\nu}\; + & \nonumber \\
 & \ds +\, \frac{1}{2} C_{\nu\rho\lambda}\left(\nabla^{\mu}K^{\rho\lambda\nu}\right)
f_{\mu} \bigg] = 0 \, , &
\end{eqnarray}
where $f_{\mu;\nu}\equiv (\nabla_{\nu}f_{\mu})_{x=z}\,$,
$f_{\mu;\nu\rho} \equiv (\nabla_{\rho}\nabla_{\nu} f_{\mu})_{x=z}\,$, and the
shorthand notation
$$
D^{\mu\nu} \equiv K^{[\mu}{}_{\lambda\rho} C^{\rho\lambda\nu]} +
\frac{1}{2}K_{\lambda\rho}{}^{[\mu} C^{\nu]\rho\lambda}
$$
is introduced for later convenience. Owing to the arbitrariness of the
function $f_{\mu}(x)$, the terms proportional to its independent derivatives
separately vanish. The independent derivatives are found by making use of
the decomposition into components orthogonal and parallel to the world-sheet
$x^{\mu}=z^{\mu}(\xi)$:
\begin{subequations} \label{jna21}
\begin{equation} \label{jna21a}
f_{\mu;\lambda} = f^{\orto}_{\mu\lambda} + u^a_{\lambda} \nabla_a f_{\mu} \, ,
\end{equation}
\begin{equation} \label{jna21b}
f_{\mu;(\lambda\rho)} = f^{\orto}_{\mu\lambda\rho} + 2 f^{\orto}_{\mu(\lambda
a} u^a_{\rho)} + f_{\mu ab} u^a_{\lambda} u^b_{\rho} \, ,
\end{equation}
\begin{equation} \label{jna21c}
f_{\mu;[\lambda\rho]}=\frac{1}{2}{R^{\sigma}}_{\mu\lambda\rho}f_{\sigma}\,.
\end{equation}
\end{subequations}
Here, the orthogonal and parallel components are obtained by using the
projectors
\begin{equation} \label{jna22}
\Pn^{\mu}_{\nu} = \delta^{\mu}_{\nu} - u_a^{\mu}u^a_{\nu}, \qquad
\Pp^{\mu}_{\nu} = u_a^{\mu} u^a_{\nu}  \, .
\end{equation}
More precisely, $f^{\orto}_{\mu\lambda} = \Pn^{\sigma}_{\lambda}
f_{\mu;\sigma}$,
$f^{\orto}_{\mu\lambda\rho} = \Pn^{\sigma}_{\lambda} \Pn^{\nu}_{\rho}
f_{\mu;\sigma\nu}$, $f^{\orto}_{\mu\lambda a} = \Pn^{\sigma}_{\lambda}
u_a^{\nu} f_{\mu;(\sigma\nu)}$ and $f_{\mu ab} = u_a^{\sigma} u_b^{\nu}
f_{\mu;(\sigma\nu)}$. Direct calculation yields
\begin{equation} \label{jna23}
\begin{array}{lcl}
f_{\mu ab} & = & \nabla_{(a} \nabla_{b)} f_{\mu} - (\nabla_a u_b^{\nu})
f^{\orto}_{\mu\nu} \, , \\
f^{\orto}_{\mu\rho a} & = & \ds \Pn^{\nu}_{\rho} \nabla_a f^{\orto}_{\mu\nu} +
(\nabla_a u^b_{\rho}) \nabla_b f_{\mu} + \frac{1}{2} \Pn^{\lambda}_{\rho}
u_a^{\nu} {R^{\sigma}}_{\mu\nu\lambda} f_{\sigma} \, , \\
\end{array}
\end{equation}
which tells us that the only independent components on the surface
$x^{\mu}=z^{\mu}(\xi)$ are $f_{\mu}$, $f^{\orto}_{\mu\nu}$ and
$f^{\orto}_{\mu\nu\rho}$. We can now use (\ref{jna21}) and (\ref{jna23})
in the equations (\ref{jna20}) to group the
coefficients into terms proportional to the independent derivatives of
$f_{\mu}$. The obtained equation has the following general structure:
\begin{eqnarray}
& \ds \int d^{p+1}\xi \sqrt{-\gamma}\, \Big[ X^{\mu\nu\rho}
f^{\orto}_{\mu\nu\rho} + X^{\mu\nu} f^{\orto}_{\mu\nu} + X^{\mu} f_{\mu} + &
\nonumber \\
& \ds  +\nabla_a\left( X^{\mu\nu a} f^{\orto}_{\mu\nu} + X^{\mu ab}
\nabla_b f_{\mu} + X^{\mu a} f_{\mu} \right)  \Big] =0 \,, & \nonumber
\end{eqnarray}
where $X$ terms are composed of various combinations of multipole
coefficients $B^{\mu\nu}$, $B^{\mu\nu\rho}$ and $C^{\lambda\mu\nu}$,
external fields $K^{\lambda}{}_{\mu\nu}$ and $R^{\mu}{}_{\nu\rho\sigma}$,
and their derivatives. In all the expressions, the external fields are
evaluated on the surface $x^{\mu}=z^{\mu}(\xi)$. Owing to the fact that
$f_{\mu}$, $f^{\orto}_{\mu\nu}$ and $f^{\orto}_{\mu\nu\rho}$ are independent
functions on the world-sheet, we deduce that the first three $X$ terms must
separately vanish.

The equation $X^{\mu\nu\rho}=0$ has a simple algebraic form
\begin{equation} \label{SpinskaJnaXtri}
\Pn^{(\nu}_{\lambda} \Pn^{\sigma)}_{\rho} B^{\mu\lambda\rho} = 0 \,.
\end{equation}
Its general solution is
\begin{equation} \label{RazlaganjeB}
B^{\mu\nu\rho} = 2\,u_a^{(\mu} J^{\nu)\rho a} + N^{\mu\nu a} u_a^{\rho}\,,
\end{equation}
where the new coefficients $J^{\mu\nu a}$ and $N^{\mu\nu a}$ are subject to
the algebraic constraints
$$
J^{\mu\nu [a}u_{\nu}^{b]}=0\,, \qquad
J^{\mu\nu a} = - J^{\nu\mu a}\,,\qquad
N^{\mu\nu a} = N^{\nu\mu a}\,.
$$
In what follows, we shall see that the coefficients $N^{\mu\nu a}$ drop from
the world-sheet equations, while $J^{\mu\nu a}$ currents couple to the
background curvature.

The equations $X^{\mu\nu}=0$ and $X^{\mu}=0$ are much more complicated. The
procedure goes as follows. First, we use the above decomposition of
$B^{\mu\nu\rho}$ to perform a similar split of the $B^{\mu\nu}$
coefficients. As a result, new free parameters $m^{ab}(\xi)$ appear to
characterize the leading term of the coefficient $B^{\mu\nu}$. Then, the
equations $X^{\mu\nu}=0$ and $X^{\mu}=0$ are rewritten in terms of the
undetermined parameters $m^{ab}$, $J^{\mu\nu a}$ and $C^{\lambda\mu\nu}$,
and properly rearranged. The coefficients $N^{\mu\nu a}$ turn out to
completely disappear. Skipping the details of the diagonalization procedure,
which has thoroughly been demonstrated in Ref. \cite{Vasilic2007}, we
display the resulting world-sheet equations:
\begin{widetext}
\begin{subequations} \label{WorldSheet}
\begin{equation} \label{JnaPrecesije}
\Pn^{\mu}_{\lambda}\Pn^{\nu}_{\rho} \left( \nabla_a J^{\lambda\rho a} +
D^{\lambda\rho} \right) = 0 \,,
\end{equation}
\begin{equation} \label{JnaKretanja}
  \nabla_b \Big[m^{ab}u_a^{\mu}-2u^b_{\lambda}\left(\nabla_a J^{\mu\lambda a} +
   D^{\mu\lambda} \right) + u_c^{\mu}u^c_{\rho} u^b_{\lambda}
   \left( \nabla_a J^{\rho\lambda a} + D^{\rho\lambda}\right)\Big] =
  \,u_a^{\nu} J^{\lambda\rho a} R^{\mu}{}_{\nu\lambda\rho} +
   \frac{1}{2} C_{\nu\rho\lambda} \nabla^{\mu} K^{\rho\lambda\nu}.
\end{equation}
\end{subequations}
\end{widetext}
The world-sheet equations (\ref{WorldSheet}) describe the dynamics of a thick
$p$-brane in $D$-dimensional spacetime with curvature and torsion. The
coefficients $m^{ab}$, $J^{\mu\nu a}$ and $C^{\lambda\mu\nu}$ are free parameters
of the theory. While $m^{ab}$ represents the effective stress-energy tensor of
the brane, the $J^{\mu\nu a}$ and $C^{\lambda\mu\nu}$ currents are related to its
total internal angular momentum and spin, respectively. In the particle case, our
world-line equations agree with the known results in literature
\cite{Trautman1972, Hehl1976a, Yasskin1980, Nomura1991, Nomura1992}.

Having solved the equations $X^{\mu\nu\rho}=X^{\mu\nu}=X^{\mu}=0$, we are left
with the surface integral that vanishes itself:
\begin{equation} \label{jna25}
\int_{\del\cM} d^p\lambda \sqrt{-\h} n_a \left( X^{\mu\nu a}
f^{\orto}_{\mu\nu} + X^{\mu ab}\nabla_b f_{\mu} +
X^{\mu a} f_{\mu} \right) = 0 \,.
\end{equation}
The components $f^{\orto}_{\mu\nu}$ and $f_{\mu}$, when  evaluated on the
boundary, are mutually independent, but $\nabla_a f_{\mu}$ is not. This is
why we decompose the $\nabla_a$ derivative into components orthogonal and
parallel to the boundary:
\begin{equation} \label{jna26}
\nabla_a f_{\mu} = n_a \nabla_{\orto} f_{\mu} + v^i_a \nabla_i f_{\mu} \,.
\end{equation}
Here, $\nabla_{\orto}\equiv n^a\nabla_a$, $\nabla_i$ is the total covariant
derivative on $\del\cM$, and $v^a_i$ are the boundary coordinate vectors (see
the Appendix for details). Now, $f^{\orto}_{\mu\nu}$, $\nabla_{\orto} f_{\mu}$
and $f_{\mu}$ are mutually independent, and the equation (\ref{jna25}) yields
three sets of boundary conditions:
\begin{widetext}
\begin{subequations} \label{Boundary}
\begin{equation} \label{Boundary1}
J^{\mu\nu a}n_a n_{\nu} \Big|_{\del\cM}=0\,, \qquad
\Pn^{\mu}_{\lambda}\Pn^{\nu}_{\rho} J^{\lambda\rho a} n_a \Big|_{\del\cM} =0\,,
\end{equation}
\begin{equation} \label{Boundary2}
  n_b \Big[ m^{ab}u_a^{\mu} -
  2 u^b_{\rho}\left( \nabla_a J^{\mu\rho a} + D^{\mu\rho} \right)
  +\, u_c^{\mu}u^c_{\sigma} u^b_{\nu} \left( \nabla_a
  J^{\sigma\nu a} + D^{\sigma\nu} \right) \Big] \Big|_{\del\cM}
  =\nabla_i \left( N^{ij}v_j^{\mu} +
  2 J^{\mu\nu a} n_a v^i_{\nu} \right)  \Big|_{\del\cM}  \,.
\end{equation}
\end{subequations}
\end{widetext}
The new free parameters $N^{ij} \equiv N^{\mu\nu a} n_a v^i_{\mu} v^j_{\nu}$
are defined on the boundary, and appear nowhere else. This situation is
familiar from the analysis of thick branes in torsionless spacetimes
\cite{Vasilic2007}. In fact, our world-sheet equations and boundary
conditions reduce to those of Ref. \cite{Vasilic2007} in the limit of
vanishing torsion.

\subsection{\label{Interpretation}Interpretation}

The world-sheet equations (\ref{WorldSheet}), and boundary conditions
(\ref{Boundary}) describe the dynamics of a thick brane in Riemann-Cartan
spacetime. The free coefficients $m^{ab}$, $J^{\mu\nu a}$,
$C^{\lambda\mu\nu}$ and $N^{ij}$ characterize the internal structure of the
brane. The tensor $m^{ab}$ represents the effective $(p+1)$-dimensional
stress-energy of the brane interior, while $N^{ij}$ stands for the
stress-energy of the brane boundary. The coefficients $J^{\mu\nu a}$ are the
world-sheet currents associated with the total internal angular momentum of
the brane, and $C^{\lambda\mu\nu}$ is the spin-tensor of the constituent
matter.

The dynamical equations that we have obtained differ from those of Ref.
\cite{Vasilic2007} by the presence of the spin-torsion couplings. By
inspecting their form, we realize that branes made of scalar matter can not
probe spacetime torsion. This is a generalization of the known result
concerning thick particles \cite{Trautman1972, Hehl1976a, Yasskin1980,
Nomura1991, Nomura1992}. We emphasize though that our predictions concerning
zero-size particles disagree with the existing literature. In the next
section, we shall demonstrate this in a simple example.

\subsubsection{Symmetries}

The symmetries of the world-sheet equations
(\ref{WorldSheet}) and boundary conditions (\ref{Boundary}) are basically
the same as in the case of spinless matter \cite{Vasilic2007}. It is only
that, in addition to $B^{\mu\nu}=\cO_0$, $B^{\mu\nu\rho}=\cO_1$, the
condition $C^{\lambda\mu\nu}=\cO_1$ is needed to ensure the existence of
extra symmetry 2. The transformation law of the free parameters $m^{ab}$,
$J^{\mu\nu a}$, $C^{\lambda\mu\nu}$ and $N^{ij}$ is obtained from the known
symmetry properties of the $B$ and $C$ coefficients. With respect to
diffeomorphisms, the coefficients
\begin{itemize}
\item $m^{ab}$, $J^{\mu\nu a}$, $C^{\lambda\mu\nu}$ and $N^{ij}$ are
      tensors of the type defined by their index structure.
\end{itemize}
Thus, $m^{ab}$ is a world-sheet tensor, $J^{\mu\nu a}$ is a spacetime tensor
and a world-sheet vector, $C^{\lambda\mu\nu}$ is a spacetime tensor, and
$N^{ij}$ is a boundary tensor. The coefficients which lack certain type of
indices transform as scalars under corresponding reparametrizations. For
example, $J^{\mu\nu a}$ is a scalar with respect to the boundary
reparametrizations, while $N^{ij}$ is a scalar with respect to spacetime and
world-sheet diffeomorphisms.

The \emph{extra symmetry} 1 is an algebraic symmetry, which ensures that
only gauge invariant coefficients appear in properly diagonalized world-sheet equations. Using (\ref{extra1}) and (\ref{extraC}), we indeed find
that our free parameters transform trivially:
\begin{equation} \label{jna38}
\delta_1 m^{ab} =
\delta_1 J^{\mu\nu a} =
\delta_1 C^{\lambda\mu\nu} =
\delta_1 N^{ij} = 0 \,.
\end{equation}
It has been shown in Ref. \cite{Vasilic2007} that the peculiar $N^{ij}$
coefficients that live exclusively on the boundary are a consequence of the
constraint (\ref{extra1b}) that parameters of the extra symmetry 1 obey. If
not for this, the transformation law $\delta_1 N^{\mu\nu a} =
\epsilon^{\mu\nu a}$ would imply that $N^{\mu\nu a}$ are pure gauge
\emph{everywhere}. Physically, the $N^{ij}$ coefficients represent a
correction to the effective $p$-dimensional stress-energy tensor of the
brane boundary, very much like $m^{ab}$ is $(p+1)$-dimensional effective
stress-energy tensor of the brane itself. The best way to see this is to
consider a brane with extra massive matter attached to its boundary. The
procedure has thoroughly been demonstrated in Ref. \cite{Vasilic2007}, where
an infinitely thin string with massive thick particles attached to its ends
has been considered.

The \emph{extra symmetry} 2 has been defined in Sec.\ \ref{Formalism} as the
symmetry generated by the change of the surface $x^{\mu}=z^{\mu}(\xi)$ used
in the $\delta$-function expansion. The transformation laws (\ref{extra2})
and (\ref{extraC}), thus obtained, can be rewritten in terms of the free
coefficients $m^{ab}$, $J^{\mu\nu a}$, $C^{\lambda\mu\nu}$ and $N^{ij}$.
Using the decomposition of the parameters $\epsilon^{\mu}$ into components
orthogonal and parallel to the world-sheet,
$\epsilon^{\mu}=\epsilon^{\mu}_{\orto}+u_a^{\mu}\epsilon^a$, we find:
\begin{widetext}
\begin{subequations} \label{jna40}
\begin{equation} \label{jna40d}
\delta_2 z^{\mu} = \epsilon^{\mu}_{\orto} + u_a^{\mu}\epsilon^a  \,,
\end{equation}
\begin{equation} \label{jna40a}
\delta_2 m^{ab} = - \left( u^c_{\mu} m^{ab} + u^{(a}_{\mu} m^{b)c} \right)
\nabla_c \epsilon^{\mu}_{\orto} - m^{bc} \nabla_c \epsilon^a -
 m^{ac} \nabla_c \epsilon^b + \epsilon^c\nabla_c m^{ab} \,,
\end{equation}
\begin{equation} \label{jna40c}
\delta_2 J^{\mu\nu a} = - m^{ab} u_b^{[\mu}\epsilon^{\nu]}_{\orto} \,,\qquad
\delta_2 N^{ij} = - m^{ab} v^i_a v^j_b \epsilon^c n_c \,,
\end{equation}
\begin{equation} \label{jna40b}
\delta_2 C^{\lambda\mu\nu} = 0 \,.
\end{equation}
\end{subequations}
\end{widetext}
The transformation laws (\ref{jna40}) are used for fixing the gauge freedom
of the world-sheet equations. As explained in Ref. \cite{Vasilic2007}, the
gauge fixing of the extra symmetry 2 corresponds to the choice of the
\emph{central surface of mass}---the surface that approximates a branelike
matter distribution. In the particle case, it coincides with the usual
notion of the centre of mass. It has been shown that an appropriate gauge
fixing ensures that particle trajectories in flat, torsionless spacetimes
coincide with straight lines. In the case of higher branes, the central
surface of mass can be chosen to eliminate the boost degrees of freedom from
the angular momentum charge densities $J^{\mu\nu 0}$. This is done by using
the transformation law (\ref{jna40c}). The residual extra symmetry 2 can
then be used to fix the trace of the boundary stress-energy $N^{ij}$. In the
string dynamics, the boundary is one-dimensional, and there is only one $N$
coefficient, which can, therefore, be completely gauged away.

\subsubsection{Single-pole limit}

The important limit of an infinitely
thin brane is obtained by discarding dipole terms in the multipole expansion
(\ref{PoleDipole}). The resultant expression contains no $\delta$-function
derivatives, and is called single-pole approximation. In our case, this is
achieved by imposing the constraint $B^{\mu\nu\rho}=0$.

The consequences of the new constraint are far reaching. We can immediately
write the single-pole equations by using the relation (\ref{RazlaganjeB})
that establishes 1--1 correspondence between the coefficients
$B^{\mu\nu\rho}$ and the world-sheet currents $J^{\mu\nu a}$ and $N^{\mu\nu
a}$. Considering the fact that $N^{ij}$ are the only surviving components of
the coefficient $N^{\mu\nu a}$, the constraint $B^{\mu\nu\rho}=0$ is
rewritten as
\begin{equation}\label{SinglePoleConstraint}
J^{\mu\nu a}=0\,,\qquad N^{ij}=0\,.
\end{equation}
Substituting (\ref{SinglePoleConstraint}) into (\ref{WorldSheet}),
(\ref{Boundary}), we obtain the single-pole world-sheet equations
\begin{subequations} \label{WorldSheetSingle}
\begin{equation} \label{JnaPrecesijeSingle}
\Pn^{\mu}_{\lambda}\Pn^{\nu}_{\rho} D^{\lambda\rho} = 0 \,,
\end{equation}
\begin{equation} \label{JnaKretanjaSingle}
\nabla_b \Big(m^{ab}u_a^{\mu}-2u^b_{\lambda}D^{\mu\lambda} +
u_c^{\mu}u^c_{\rho} u^b_{\lambda}D^{\rho\lambda}\Big) =
\frac{1}{2} C_{\nu\rho\lambda} \nabla^{\mu} K^{\rho\lambda\nu},
\end{equation}
\end{subequations}
and the single-pole boundary conditions
\begin{equation} \label{BoundarySingle}
n_b \Big( m^{ab}u_a^{\mu} -
 2 u^b_{\rho}D^{\mu\rho} +
u_c^{\mu}u^c_{\sigma} u^b_{\nu} D^{\sigma\nu}\Big)\Big|_{\del\cM}=0 \,.
\end{equation}
The equations (\ref{WorldSheetSingle}) and (\ref{BoundarySingle}) describe
the motion of an infinitely thin brane in Riemann-Cartan spacetime. Compared
to the pole-dipole approximation, there are two striking differences. First,
the coupling of the spacetime curvature to the internal angular momentum of
the brane is missing. This is something one would expect to hold for
spinless matter, only. Indeed, if the brane has no thickness, the
transversal internal motion is not possible, and the internal orbital
angular momentum vanishes \cite{Vasilic2007}. The spin part of the total
angular momentum, however, is expected to survive. Thus, the vanishing of
the spin-curvature coupling in the single-pole limit comes as a surprise.

Another striking consequence of the single-pole limit is the algebraic
nature of the precession equations (\ref{JnaPrecesijeSingle}). This new
constraint restrains the allowed forms of the spin-tensor in the presence of
the background torsion. If the torsion is absent, the constraint is
identically satisfied. In the next section, this unusual behaviour will be
studied in the particle example.

\section{\label{Examples}Examples}

To illustrate the relevance of the brane dynamics in Riemann-Cartan spacetimes,
we shall discuss two important examples: zero-size particle and infinitely thin
string. In the particle case, the novel spin-curvature coupling is shown to
contradict the results of the existing literature. In the string case, the action
of Refs. \cite{B1, B2, B3, B4} is recovered, and the Kalb-Ramond
antisymmetric field ${\cal B}^{\mu\nu}$ given a geometric interpretation.

\subsection{\label{Particle}Particle}

The world-sheet of a particle is one-dimensional, and is called world-line.
We shall parametrize it with the proper distance $\tau$, thereby fixing the
reparametrization invariance:
$$
\gamma = u^{\mu}u_{\mu} =-1 \, .
$$
Here, and in what follows, the indices $a,b,\dots$ are omitted, as they take
only one value. We shall restrict to a {\it zero-size} particle in
$4$-dimensional spacetime. Thus, the single-pole equations
(\ref{WorldSheetSingle}) yield:
\begin{subequations} \label{ParticleSingle}
\begin{equation} \label{ParticleJnaKretanja}
\nabla\left(m u^{\mu}+2D^{\mu\nu}u_{\nu}\right) =
\frac{1}{2} C_{\nu\rho\lambda} \nabla^{\mu} K^{\rho\lambda\nu},
\end{equation}
\begin{equation} \label{ParticleJnaPrecesije}
\Pn^{\mu}_{\lambda}\Pn^{\nu}_{\rho} D^{\lambda\rho} = 0 \,,
\end{equation}
\end{subequations}
where $\Pn^{\mu}_{\nu}=\delta^{\mu}_{\nu}+u^{\mu}u_{\nu}$ is the orthogonal
world-line projector. As we can see, the spin couples only to contorsion,
which means that \emph{point particles follow geodesic trajectories in
torsionless spacetimes}. At the same time, the absence of torsion
trivializes the equation (\ref{ParticleJnaPrecesije}), and no information on
the behavior of the spin-vector is available. If the background torsion is
nontrivial, a geodesic deviation appears, but also a very strong constraint
on the spin-vector.

Let us now apply the obtained equations to the Dirac particle. The basic
property of Dirac matter is the total antisymmetry of its spin tensor
$\sigma^{\lambda\mu\nu}$. As a consequence, the coefficients
$C^{\lambda\mu\nu}$ are also totally antisymmetric. If we define the spin
vector $s^{\mu}$ by $C^{\mu\nu\rho} \equiv
e^{\mu\nu\rho\lambda}s_{\lambda}$, and the axial component of the contorsion
$K^{\mu}$ as $K^{\mu} \equiv e^{\mu\nu\rho\lambda} K_{\nu\rho\lambda}$,
where $e^{\mu\nu\rho\sigma}$ is the covariant totally antisymmetric
Levi-Civita tensor, the equations (\ref{ParticleSingle}) become
\begin{subequations} \label{jna6}
\begin{equation} \label{jna6a}
\nabla\left( mu^{\mu} + K^{[\mu}s^{\nu]} u_{\nu} \right) + \frac{1}{2}
s^{\nu} \nabla^{\mu}K_{\nu} =0 ,
\end{equation}
\begin{equation} \label{jna6b}
K_{\perp}^{[\mu}s_{\perp}^{\nu]} = 0.
\end{equation}
\end{subequations}
The equation (\ref{jna6b}) implies that the orthogonal component of
$s^{\mu}$ always orients itself along the background direction
$K_{\perp}^{\mu}$. This unusual behavior suggests the possibility that the
spin itself might vanish in the zero-size limit. Moreover, the same
suggestion comes from the disappearance of the total angular momentum
$J^{\mu\nu}$ in the single-pole approximation. Indeed, the zero-size limit
naturally rules out the orbital part of the angular momentum. If
$J^{\mu\nu}$ is seen as the sum of orbital and spin parts, its disappearance
inevitably implies the disappearance of the spin itself.

To justify this scenario, the authors of Refs. \cite{Vasilic2008a, Vasilic2008b}
have analyzed the wave packet solutions of the flat space Dirac equation. The
idea was to check if the wave packets could be viewed as zero-size objects. For
that purpose, the wave packet size $\ell$, and its wavelength $\lambda$ are
considered in the limit $\ell\to 0$, $\lambda / \ell \to 0$. It has been
discovered that the wave packet spin and orbital angular momentum disappear
simultaneously in this limit. Thus, the spin vector $s^{\mu}$ vanishes in the
single-pole approximation, and the particle trajectory becomes a geodesic line
even in the presence of torsion.

The single-pole results obtained in this section disagree with the results found
in literature \cite{Hehl1976a, Yasskin1980, Nomura1991}. In these early
approaches, the antisymmetric part of the stress-energy tensor $\tau^{[\mu\nu]}$
has been treated as an independent variable, in spite of the restriction
(\ref{ZakonOdrzanjaSigma}). This imposed unnecessary constraints on
$\sigma^{\lambda\mu\nu}$. In particular, the spin of the Dirac particle was ruled
out. In our approach, the only independent variables are the spin-tensor
$\sigma^{\lambda\mu\nu}$ and the symmetric Belinfante tensor $\theta^{\mu\nu}$.
As a result, our single-pole limit is less restrictive, and allows an equal
treatment of all massive elementary fields.

\subsection{\label{String}String}

The string trajectory is a two-dimensional world-sheet with one-dimensional
boundary. As in the particle case, the boundary line will be parametrized
with the proper distance $\tau$, and the indices $i,j,\dots$, which take
only one value, will be omitted. Thus, the boundary metric $\h$, and the
tangent vector $v^a$ satisfy
$$
\h = v^av_a =-1 \, .
$$
The only peculiarity of the string dynamics, as compared to higher branes,
is the possibility to gauge away the $N^{ij}$ coefficients. Indeed, there is
only one such component in the string case, and one free parameter in the
transformation law (\ref{jna40c}). After fixing the gauge $N=0$, the
$\epsilon^a$ part of the extra symmetry~2 reduces to reparametrizations.

In what follows, we shall consider {\it infinitely thin strings}, and thus, make
use of the single-pole equations (\ref{WorldSheetSingle}) and
(\ref{BoundarySingle}). In this approximation, the extra symmetries are trivial,
and the coefficients $J^{\mu\nu a}$ and $N$ are zero. The remaining coefficients
$m^{ab}$ and $C^{\rho\mu\nu}$ carry the information on the type of matter the
string is made of.

Our idea is to try and find the string type whose classical dynamics
coincides with that of Refs. \cite{B1, B2, B3, B4}. There, the string is
coupled to the Kalb-Ramond antisymmetric field ${\cal B}_{\mu\nu}$, commonly
interpreted as the torsion potential, $K_{\mu\nu\rho} \propto \nabla_{\mu}
{\cal B}_{\nu\rho} + \nabla_{\nu} {\cal B}_{\rho\mu} + \nabla_{\rho} {\cal
B}_{\mu\nu}$ \cite{G1, G2, G3, G4, G5, Nepomechie1985, Freund1982}. If we
accept this picture, however, we find that no choice of the coefficients
$m^{ab}$ and $C^{\rho\mu\nu}$ leads to the satisfactory solution. Indeed,
the spin-torsion couplings in the world-sheet equations
(\ref{WorldSheetSingle}) contain torsion derivatives, which do not exist in
the string dynamics of Refs. \cite{B1, B2, B3, B4}. The l.h.s. contains only
derivatives parallel to the world-sheet, which can not compensate for the
orthogonal derivatives on the r.h.s. The only way to get rid of these is to
have $C_{\rho\mu\nu}\propto u^a_{\rho}u^b_{\mu}u^c_{\nu}$, but the assumed
total antisymmetry of $K^{\mu\nu\rho}$ rules out this choice. Thus, whatever
choice of $C^{\rho\mu\nu}$ is made, the resulting word-sheet equations will
contain undesirable couplings to torsion derivatives. We are led to the
conclusion that the usual interpretation of ${\cal B}^{\mu\nu}$ as the
torsion potential is not supported by the classical string dynamics in
Riemann-Cartan spacetimes.

In what follows, we shall continue searching for the string type that realizes
the correct ${\cal B}^{\mu\nu}$ coupling, whatever geometric interpretation of
${\cal B}^{\mu\nu}$ field may be. Skipping the details of our pursuit, we shall
describe the scheme that we have found to work.

First, we restrict our attention to spacetimes characterized by the
contorsion of the form
\begin{equation} \label{Torzija}
K^{\mu\nu\rho} = K^{\mu\nu}K^{\rho}\,,
\end{equation}
where $K^{\mu\nu}\equiv -K^{\nu\mu}$ is an arbitrary antisymmetric tensor, and
$K^{\rho}$ is an arbitrary vector field. It is obvious that this decomposition is
not unique. Indeed, the transformation $K^{\mu\nu}\to\alpha K^{\mu\nu}$,
$K^{\rho}\to\alpha^{-1}K^{\rho}$ leaves the contorsion $K^{\mu\nu\rho}$
invariant. Using this freedom, we shall fix the norm of the vector field
$K^{\rho}$ to be
$$
K^{\rho}K_{\rho} = \kappa \,,
$$
where $\kappa$ $=$ $1$, $-1$ or $0$, depending on whether $K^{\rho}$ is
spacelike, timelike or lightlike vector. From now on, we shall assume that
$\kappa$ takes only one value in the whole spacetime.

Our second assumption concerns the spin tensor coefficients $C^{\rho\mu\nu}$. We
specify their form by the relation
\begin{equation} \label{C}
C^{\rho\mu\nu} = s K^{\rho} u^{\mu\nu} \,,
\end{equation}
where $u^{\mu\nu} \equiv e^{ab}u^{\mu}_a u^{\nu}_b$, and $s$ is a constant that
measures the spin magnitude.

With these assumptions, the world-sheet equations (\ref{WorldSheetSingle}),
and boundary conditions (\ref{BoundarySingle}) are rewritten in terms of the
free coefficients $m^{ab}$, and the external fields $K^{\mu\nu}$. First,
we calculate the $D^{\mu\nu}$ tensor, and find that it reduces to
$$
D^{\mu\nu} = s \kappa\, e^{ab} u^{[\mu}_a K^{\nu]}{}_b  \,,
$$
where $K^{\mu a}\equiv K^{\mu\nu}u^a_{\nu}\,$. The precession equations
(\ref{JnaPrecesijeSingle}) are then automatically satisfied, and we are left
with the world-sheet equations (\ref{JnaKretanjaSingle}), and boundary
conditions (\ref{BoundarySingle}). Now, we calculate the right-hand side of
(\ref{JnaKretanjaSingle}), and find
$$
C_{\nu\rho\lambda} \nabla^{\mu} K^{\rho\lambda\nu} =
s\kappa \left( F^{\mu\rho\lambda} u_{\rho\lambda} -
2 e_{ab}\nabla^a K^{b\mu} \right) \,,
$$
with
$$
F^{\mu\nu\rho} \equiv
\nabla^{\mu} K^{\nu\rho} + \nabla^{\nu} K^{\rho\mu} +
\nabla^{\rho} K^{\mu\nu} \,.
$$
With the help of these expressions, the world-sheet equations are rewritten as
\begin{equation} \label{EqnWithMbar}
\nabla_b \left(\bar m^{ab} u^{\mu}_a \right) =
\frac{s\kappa}{2}F^{\mu\rho\lambda} u_{\rho\lambda} \,,
\end{equation}
where the new parameters $\bar m^{ab}$ are related to $m^{ab}$ as follows:
$$
\bar m^{ab} \equiv m^{ab} -
\frac{s\kappa}{2}\gamma^{ab} K^{\mu\nu} u_{\mu\nu}  \,.
$$
Using the total antisymmetry of the $F^{\mu\rho\lambda}$ coefficients in
(\ref{EqnWithMbar}), the new coefficients $\bar m^{ab}$ are shown to be
covariantly conserved,
\begin{equation} \label{Conservation}
\nabla_b \bar m^{ab} = 0 \,.
\end{equation}
This means that Nambu-Goto matter is allowed as the constituent matter of
our string. Indeed, by demanding $\bar m^{ab}=T\gamma^{ab}$, where $T$ is a
constant commonly interpreted as the string tension, the condition
(\ref{Conservation}) is automatically satisfied. At the same time, we obtain
the world-sheet equations in their final form
\begin{equation} \label{EqnFinal}
\nabla_a u^{a\mu}  =
\frac{s\kappa}{2T}\, F^{\mu\rho\lambda} u_{\rho\lambda} \,.
\end{equation}
Following the same procedure, the boundary conditions are rewritten as
\begin{equation} \label{BoundaryFinal}
n^a \left( u^{\mu}_a + \frac{s\kappa}{T} K^{\mu b} e_{ab}\right)
\Big|_{\del\cM} = 0 \,.
\end{equation}
The world-sheet equations (\ref{EqnFinal}), and boundary conditions
(\ref{BoundaryFinal}) are exactly the same as obtained by varying the string
action of Refs. \cite{B1, B2, B3, B4}. This action describes a string
interacting with an additional external antisymmetric field ${\cal
B}_{\mu\nu}(x)$, and has the form
$$
S = T \int d^2\xi \sqrt{-\gamma}
\left[ g_{\mu\nu}(x) u^{\mu}_a u^{\nu}_b \gamma^{ab} +
{\cal B}_{\mu\nu}(x) u^{\mu}_a u^{\nu}_b e^{ab} \right] ,
$$
where the world-sheet metric $\gamma_{ab}$ is considered as an independent
variable. One can verify that it is indeed minimized by our equations
(\ref{EqnFinal}) and (\ref{BoundaryFinal}), provided the identification
$$
{\cal B}^{\mu\nu} \equiv \frac{s\kappa}{T} K^{\mu\nu}
$$
is made. Thus, the Kalb-Ramond antisymmetric field ${\cal B}^{\mu\nu}$
is recognized as a {\it part of the spacetime torsion}.

Note that the geometric interpretation found in literature is not the same.
There, the external field ${\cal B}^{\mu\nu}$ has commonly been treated as a
torsion potential, rather than the torsion itself. The authors of Refs.
\cite{G1, G2, G3, G4, G5} studied the influence of the string background
fields on the string dynamics, and succeeded in rewriting the world-sheet
equations in geometric terms. In this setting, the field strength of the
${\cal B}^{\mu\nu}$ field turned out to define the torsion part of the
modified geometry. We must emphasize, however, that this new geometry is
characterized by the presence two connections, and does not belong to the
class of Riemann-Cartan geometries considered in this paper.

In Ref. \cite{Nepomechie1985}, a similar line of reasoning has been applied
to the string low-energy effective action. This action governs the dynamics
of the string background fields---the spacetime metric $g_{\mu\nu}$,
Kalb-Ramond field ${\cal B}_{\mu\nu}$ and the dilaton $\Phi$. It was shown
that the field strength of the Kalb-Ramond field could be absorbed in the
antisymmetric part of a new connection. In this approach, however, the
derived torsion is free, in the sense that no string-torsion coupling is
specified. Thus, this approach is complementary to our treatment of strings
in fixed backgrounds, with no background dynamics specified.

Finally, let us mention that the same discussion applies to the pure
geometric considerations of Ref. \cite{Freund1982}.

In summary, there are different ways of relating the ${\cal
B}_{\mu\nu}$ field to torsion, and our result follows from the particular
approach of treating probe strings in Riemann-Cartan backgrounds.

\section{\label{ConclusionSection}Concluding remarks}

In this paper, we have analyzed classical dynamics of brane like objects in
backgrounds of nontrivial geometry. In particular, our target space is
characterized by both, curvature and torsion. The type of matter fields the
brane is made of has not been specified. We have only assumed that matter
fields are sharply localized around a brane.

The method we have used is a generalization of the Mathisson-Papapetrou
method for pointlike matter \cite{Mathisson1937,Papapetrou1951}. It has
already been used in Refs. \cite{Vasilic2006, Vasilic2007} for the study of
strings and higher branes in Riemannian spacetimes. In this work, we have
extended the analysis to Riemann-Cartan spacetimes.

Our exposition is summarized as follows. In section \ref{Formalism}, we have
defined the conservation law of the stress-energy and spin tensors, and
eliminated the antisymmetric part of stress-energy by noticing that it is
not an independent variable. Thereby, the conservation equations are
rewritten in terms of the spin tensor $\sigma^{\lambda\mu\nu}$ and the
symmetric Belinfante tensor $\theta^{\mu\nu}$. A brief recapitulation of the
covariant multipole formalism, and its symmetry properties has been given by
invoking the results of Ref. \cite{Vasilic2007}. Then, the pole-dipole
approximation has been defined for the independent variables, only.

In section \ref{EOMSection}, the brane world-sheet equations and boundary
conditions have been obtained in a manifestly covariant form. In the
particle case, the pole-dipole result has been shown to agree with the
results of the existing literature. The single-pole limits, however, turned
out to differ. This is a consequence of the fact that earlier approaches
incorrectly treated the antisymmetric part of the stress-energy
$\tau^{[\mu\nu]}$ as an independent variable.

In Section \ref{Examples}, we have analyzed the $0$-brane and $1$-brane
examples. In the particle case, the world-line equations have been compared
to the known pole-dipole equations \cite{Yasskin1980, Nomura1991}. They are
found to coincide in the pole-dipole approximation, but have different
single-pole limits. In the string case, the world-sheet equations are
analyzed in the limit of zero thickness. By an appropriate choice of the
spin-tensor and the background torsion, we have recovered the string action
of Refs. \cite{B1, B2, B3, B4}. In particular, the Kalb-Ramond antisymmetric
tensor field ${\cal B}^{\mu\nu}$ has been recognized as the torsion itself,
rather than its potential. The apparent contradiction with the existing
literature on the subject is illusive. There are different ways of
relating the ${\cal B}_{\mu\nu}$ field to torsion, and they are often
complementary to each other. While we treat strings coupled to fixed
Riemann-Cartan backgrounds, some authors consider dynamical backgrounds with
no string couplings \cite{Nepomechie1985, Freund1982}, or employ non
Riemann-Cartan geometries \cite{G1, G2, G3, G4, G5}.

\begin{acknowledgments}
This work is partly supported by the Serbian Ministry of Science and
Technological Development, under contract No. 141036.
One of the authors (Marko Vojinovi\'c) acknowledges hospitality at the Institute
for Nuclear Research and Nuclear Energy in Sofia (Bulgaria) during his visit as
an early stage researcher, supported by the FP6 Marie Curie Research Training
Network "Forces-Universe" MRTN-CT-2004-005104.
\end{acknowledgments}

\appendix*

\section{Differential geometry of surfaces}

In this work, we deal with the geometry of surfaces embedded in
Riemann-Cartan spacetime. Let us summarize the basic notions and relations
used throughout the paper.

We shall consider a $D$-dimensional Riemann-Cartan spacetime parametrized by
the coordinates $x^{\mu}$. Its metric tensor is denoted by $g_{\mu\nu}(x)$,
and is assumed to have Minkowski signature. Given the metric, one defines
the Levi-Civita connection
$$
\cfl{\mu}{\rho\sigma} \equiv
\frac{1}{2}\,g^{\mu\lambda}
\left(\del_{\rho}g_{\lambda\sigma} +
\del_{\sigma} g_{\lambda\rho} -
\del_{\lambda} g_{\rho\sigma} \right) \, ,
$$
and the Riemannian covariant derivative $\nabla_{\lambda}$:
$$
\nabla_{\lambda} V^{\mu} \equiv \del_{\lambda} V^{\mu} +
\cfl{\mu}{\rho\lambda} V^{\rho} \, .
$$
We now introduce a $(p+1)$-dimensional surface $\cM$ parametrized by the
coordinates $\xi^a$. If the surface equation is $x^{\mu}=z^{\mu}(\xi)$, one
can define the coordinate vectors
$$
u_a^{\mu} \equiv \frac{\del z^{\mu}}{\del\xi^a} \, ,
$$
and the induced metric tensor
$$
\gamma_{ab} \equiv g_{\mu\nu}(z) u_a^{\mu} u_b^{\nu} \, .
$$
The surface $\cM$ is assumed to be everywhere regular, and the coordinates
$x^\mu$ and $\xi^a$ well defined. If it is future directed, the induced
metric has Minkowski signature. We can now define the \emph{total}
Riemannian covariant derivative $\nabla_a$, that acts on both types of
indices:
\begin{equation}\label{Al}
\nabla_a V^{\mu b} = \del_a V^{\mu b} +
\cfl{\mu}{\lambda\rho} u_a^{\rho} V^{\lambda b} +
\cfl{b}{ca} V^{\mu c} \, .
\end{equation}
Here, $\cfl{a}{bc}$ is the Levi-Civita connection on the surface, so that
the metricity conditions $\nabla_a g_{\mu\nu}=\nabla_a \gamma_{bc}=0$ are
identically satisfied.

A spacetime vector $V^{\mu}$ can uniquely be split into vectors orthogonal
and tangential to the surface by using the projectors
$$
\Pp^{\mu}_{\nu} \equiv u_a^{\mu}u^a_{\nu} \, , \qquad \Pn^{\mu}_{\nu} \equiv
\delta^{\mu}_{\nu} - u_a^{\mu}u^a_{\nu} \, .
$$
Thus, $V^{\mu} = V^{\mu}_{\orto} + V^{\mu}_{\para}$, where
$V^{\mu}_{\orto}\equiv\Pn^{\mu}_{\nu}V^{\nu}$, and $V^{\mu}_{\para} \equiv
\Pp^{\mu}_{\nu}V^{\nu}$.

The surface $\cM$ may have a boundary $\del\cM$, and we denote its
coordinates by $\lambda^i$. The boundary is assumed to satisfy the analogous
geometric requirements as the surface itself. Given the boundary $\xi^a =
\zeta^a(\lambda)$, one introduces its coordinate vectors
$$
v^a_i \equiv \frac{\del\zeta^a}{\del\lambda^i} \, ,
$$
and the induced metric
$$
\h_{ij} = \gamma_{ab}(\zeta) v^a_i v^b_j \, .
$$
The boundary connection is defined to be the Levi-Civita connection
$\cfl{i}{jk}$, so that the total covariant derivative  $\nabla_i$, which acts as
\begin{equation}\label{A2}
\nabla_i V^{\mu bj} = \del_i V^{\mu bj} +
\cfl{\mu}{\lambda\rho} v_i^{\rho} V^{\lambda bj} +
\cfl{b}{ca} v^a_i V^{\mu cj} + \cfl{j}{ki} V^{\mu bk} \, ,
\end{equation}
satisfies the metricity conditions $\nabla_i g_{\mu\nu} = \nabla_i
\gamma_{ab} = \nabla_i \h_{jk} =0$. Here, $v_i^{\mu} \equiv u_a^{\mu} v^a_i$
are the spacetime components of the boundary coordinate vectors. The
boundary projectors are defined as $\Ppp^{\mu}_{\nu}\equiv
v_i^{\mu}v^i_{\nu}$ and $\Pnn^{\mu}_{\nu}\equiv \delta^{\mu}_{\nu}-
v_i^{\mu}v^i_{\nu}$.

Throughout the paper, the covariant form of the Stokes theorem is used:
$$
\int_{\cM} d^{p+1}\xi \sqrt{-\gamma} \nabla_a V^a = \int_{\del\cM} d^p\lambda
\sqrt{-\h} n_a V^a  \, .
$$
Here, $n_a$ is the normal to the boundary. It is defined as
\begin{equation}\label{A3}
n_a = \frac{1}{p!} e_{ab_1\dots b_p} e^{i_1\dots i_p} v^{b_1}_{i_1}\dots
v^{b_p}_{i_p} \, ,
\end{equation}
where $e_{ab_1\dots b_p}$ and $e^{i_1 \dots i_p}$ are totally
antisymmetric world tensors on the surface and the boundary,
respectively. They are defined using the Levi-Civita symbols
$\lc_{ab_1\dots b_p}$ and $\lc^{i_1\dots i_p}$, and corresponding
metric determinants:
$$
e_{ab_1\dots b_p}(\xi) \equiv \sqrt{-\gamma}\lc_{ab_1\dots b_p} ,\qquad
e^{i_1 \dots i_p}(\lambda) \equiv \frac{1}{\sqrt{-\h}} \lc^{i_1\dots i_p} \, .
$$
The normal $n_a$ is always spacelike, and satisfies the following identities:
$$
n_an^a= 1 \, , \qquad n_av^a_i=0 \, , \qquad \Pn^{\mu}_{\nu} =
\Pnn^{\mu}_{\nu}
-
n^{\mu}n_{\nu} \, ,
$$
where $n^{\mu}\equiv u_a^{\mu}n^a$.

\end{document}